\documentclass[prd,aps,showpacs,preprintnumbers,11pt]{revtex4}
\usepackage{epsfig}
\usepackage{subfigure}
\begin{document}  
%\draft  
\newcommand{\be}{\begin{equation}}\newcommand{\ee}{\end{equation}}
\newcommand{\bea}{\begin{eqnarray}}\newcommand{\eea}{\end{eqnarray}}
\newcommand{\bc}{\begin{center}}\newcommand{\ec}{\end{center}}
\def\no{\nonumber}
\def\eq#1{Eq. (\ref{#1})}\def\eqeq#1#2{Eqs. (\ref{#1}) and  (\ref{#2})}
%%%%%%%%%%%%%%%%%%%%%%%%%%%%%%%%%%%%%%%%%%%%%%%%%%%
\def\lsim{\raise0.3ex\hbox{$\;<$\kern-0.75em\raise-1.1ex\hbox{$\sim\;$}}}
\def\gsim{\raise0.3ex\hbox{$\;>$\kern-0.75em\raise-1.1ex\hbox{$\sim\;$}}}
\def\slash#1{\ooalign{\hfil/\hfil\crcr$#1$}}
\def\eff{\mbox{\tiny{eff}}}
\def\order#1{{\mathcal{O}}(#1)}
\def\pppm{B^0\to\pi^+\pi^-}
\def\pzpz{B^0\to\pi^0\pi^0}
\def\pppz{B^0\to\pi^+\pi^0}
%%%%%%%%%%%%%%%%%%%%%%%%%%%%%%%%%%%%%%%%%%%%%%%%%%%
\preprint{}
\title{Test of $SU(3)$ Symmetry in Hyperon Semileptonic Decays}
\author{T. N. Pham }
\affiliation{
Centre de Physique Th\'{e}orique, CNRS,
Ecole Polytechnique, 91128 Palaiseau, Cedex, France}

\date{\today}
\begin{abstract}
Existing analyzes  of baryon semileptonic decays indicate the presence
of a small $SU(3)$ symmetry breaking in hyperon semileptonic decays, but 
to provide evidence for $SU(3)$ symmetry breaking, one would need a relation
similar to the Gell-Mann Okubo(GMO) baryon mass formula which is satisfied
to a few percents, showing evidence for $SU(3)$ symmetry breaking
in  the divergence of the vector current matrix element. In this paper,
we shall present  a similar GMO relation  for the hyperon semileptonic
decay axial vector form factors. Using these  relations and 
the measured axial vector current to vector current form factor 
ratios, we show that  $SU(3)$ symmetry breaking in hyperon
 semileptonic decays is of $5-11\%$.

\end{abstract}
\maketitle
%%%%%%%%%%%%%%%%%%%%%%%%%%%%%%%%%%%%%%%%%%%%%%%%%%%

%%%%%%%%%%%%%%%%%%%%%%%%%%%%%%%%%%%%%%%%%%%%%%%%%%%
\section{Introduction}
 The success of the Gell-Mann Okubo(GMO) mass formula shows that $SU(3)$
is a good symmetry for strong interactions. This approximate symmetry
can be incorporated into a QCD Lagrangian with $ m_{u},m_{d}\ll m_{s}$,
with $m_{s}\ll \Lambda_{\rm QCD}$,  we have an almost $SU(3)$-symmetric
Lagrangian. At low energies, an
effective chiral Lagrangian can be constructed with baryons coupled to
the pseudo-scalar meson octet, $\pi, K,\eta$ via a covariant derivative 
constructed with the derivative of the the pseudo-scalar meson field 
operator. This gives us the Goldberger-Treiman relation  for the
pion-nucleon coupling constant. This Lagrangian contains the axial vector 
current matrix elements and produces the axial vector form factors
measured in baryon semileptonic decays. At zero order in $m_{s}$,
the axial vector current form factors and the pseudo-scalar baryon
couplings are $SU(3)$-symmetric and are completely given by the two parameters
$F$ and $D$ of the $F$(antisymmetric) and $D$(symmetric) type 
coupling\cite{Donoghue}. The success of the GMO formula which can be
derived from this effective Lagrangian suggests that semileptonic
hyperon decays can also be well described by the two $SU(3)$-symmetric 
 $F$ and $D$ parameters as in the Cabibbo model \cite{Cabibbo}
for which the agreement with experiments is quite good \cite{Marshak}.
In general one expects some small $SU(3)$ symmetry breaking 
for the divergence of the vector current matrix element
 and the hyperon semileptonic decay axial vector current
 matrix elements which, unlike the vector current, are not protected 
by the Ademollo-Gatto theorem \cite{Ademollo}. Using the precise 
measured axial vector to vector form factor ratio  $g_{1}/f_{1}$ for
hyperon semileptonic decays \cite{PDG}, recent 
analyzes \cite{Ratcliffe,Song,Yamanishi} indicate the presence of a 
small $SU(3)$ symmetry breaking  in hyperon semileptonic decays. However
to have a test of $SU(3)$ symmetry, one need a relation similar to the
GMO baryon mass formula which can be written as \cite{Marshak}~:
\be
 (3/4)\Delta M +  (1/4)\Delta M^{'} =  (1/4)\Delta M^{''} + 
 (3/4)\Delta M^{'''}
\label{GMO}
\ee
with $\Delta M = m_{\Lambda} - m_{N} $, $\Delta M^{'} = m_{\Sigma} - m_{N} $,
$ \Delta M^{''}= m_{\Xi} - m_{\Sigma} $ and
 $\Delta M^{'''}= m_{\Xi} - m_{\Lambda} $.
Numerically, the l.h.s of Eq. (\ref{GMO}) is $0.1966\,\rm GeV$ while
 the r.h.s is $0.1867\,\rm GeV$ showing evidence for $SU(3)$ breaking 
for the divergence of the vector current matrix elements which in fact, 
gives the above GMO formula  by equating the matrix elements of the 
divergence of the $\Delta S=1$ vector current $\bar{u}\,\gamma_{\mu}\,s$
within the $V=1$ multiplet. The baryon mass difference is given by
 $m_{s}<B^{'}|\bar{u}\,s|B>$, with $\bar{u}\,s$ a V-spin $V=1$ scalar 
current in $SU(3)$ space. In  the limit of 
neglecting the light current quark mass $m_{u,d}$,  the  l.h.s and 
the r.h.s of Eq. (\ref{GMO}) are two  $<V=1,V_{3}=0|\bar{u}s|V=1,V_{3}=1>$ and
$<V=1,V_{3}=-1|\bar{u}s|V=1,V_{3}=0>$ matrix elements of the $V=1$ V-spin
multiplet and would be equal but opposite in sign in the limit of
 $SU(3)$ symmetry. Similarly, in the limit of $SU(3)$ symmetry,   
for the  axial vector current  matrix elements,  we have the equality of 
 $<V=1,V_{3}=-1|\bar{u}\,\gamma_{\mu}\gamma_{5}\,s|V=1,V_{3}=0>$
and $-<V=1,V_{3}=0|\bar{u}\,\gamma_{\mu}\gamma_{5}\,s|V=1,V_{3}=1>$, hence
the GMO type relation for the axial vector current form factors in   
hyperon semileptonic decays. Another non-trivial relation for 
hyperon semileptonic decays is obtained from the equality of 
two matrix elements
$<V=1,V_{3}=-1|\bar{u}\,\gamma_{\mu}\gamma_{5}\,s|V=0,V_{3}=0>$ and 
$<V=0,V_{3}=0|\bar{u}\,\gamma_{\mu}\gamma_{5}\,s|V=1,V_{3}=1>$ . In the
 following we will present test of $SU(3)$ symmetry in 
semileptonic hyperon decays and an analysis of $SU(3)$ symmetry
breaking using these relations. We  show that the amount of
 $SU(3)$ symmetry breaking in hyperon semileptonic decays is of 
$5-11\%$. 

\section{Test of $SU(3)$ Symmetry in Hyperon Semileptonic Decays}

The traditional method to obtain the GMO mass formula is to assume that
the $SU(3)$ symmetry breaking mass term in the baryon Lagrangian
transforms like the 8-th component of an $SU(3)$ octet. Nowsaday, we know
that in the standard model, $SU(3)$ symmetry breaking is given by the
current quark mass term in the QCD Lagrangian with $m_{u,d} \ll m_{s}$.
Instead of working with the quark mass term, we could obtain the GMO
relation by considering the divergence of the $\Delta S=1$  V-spin $V=1$ 
vector current $\bar{u}\,\gamma_{\mu}\,s$  or
 the U-spin $U=1$ vector current $\bar{d}\,\gamma_{\mu}\,s$ 
(putting $m_{u,d}=0$ and neglecting
isospin breaking. Consider the divergence the $V=1$ $\bar{u}\,\gamma_{\mu}\,s$ 
vector current, we have:
\be
\partial_{\mu}(\bar{u}\,\gamma_{\mu}\,s) = -i\,m_{s}\bar{u}\,s
\label{div}
\ee
Taking the matrix element of Eq. (\ref{div}) between the baryons within
a $V=1$ multiplet, we see that the baryon mass difference 
is given by the $\bar{u}\,s $ scalar current form factor at the momentum
transfer $q=0$. Since the vector current form factor on the l.h.s 
has no first order $SU(3)$ breaking according to the Ademollo-Gatto
theorem, in the limit of $SU(3)$ symmetry, 
the matrix element of $\bar{u}\,s $, like the I-spin symmetry
 for the matrix element of $\bar{u}\,d $, satisfies 
the V-spin symmetry relations  from which one obtains the GMO relation. 
There could be first order  $SU(3)$ symmetry breaking in the matrix
 element of $\bar{u}\,s $ so  
there would be  violation  to GMO mass formula. In  the limit
of $SU(3)$ symmetry, we have:
\bea
&& <\frac{1}{2}\Sigma^{0} +\frac{\sqrt{3}}{2}\Lambda|\bar{u}\,s|\Xi^{-}>= 
-<p|\bar{u}\,s|\frac{1}{2}\Sigma^{0} +\frac{\sqrt{3}}{2}\Lambda>\label{su1}\\
&& <\frac{\sqrt{3}}{2}\Sigma^{0} -\frac{1}{2}\Lambda|\bar{u}\,s|\Xi^{-}>= 
<p|\bar{u}\,s|\frac{\sqrt{3}}{2}\Sigma^{0} -\frac{1}{2}\Lambda>
\label{su2}
\eea
where  $|\Xi^{-}>=|V=1,V_{3}=1>$ ,  $|p>=|V=1,V_{3}=-1>$ ,
  $|\frac{1}{2}\Sigma^{0} +\frac{\sqrt{3}}{2}\Lambda>=|V=1,V_{3}=0>$
and $|\frac{\sqrt{3}}{2}\Sigma^{0} -\frac{1}{2}\Lambda>=|V=0,V_{3}=0>$. 
Eq. (\ref{su1}) and Eq. (\ref{su2}) are  the rotated 
V-spin version  of the two I-spin relations for the   $\bar{u}\,d$ 
matrix elements:
\bea
&& <\Sigma^{0}|\bar{u}\,d|\Sigma^{+}>=
-<\Sigma^{-}|\bar{u}\,d|\Sigma^{0}>\label{du1}\\
&& <\Lambda|\bar{u}\,d|\Sigma^{+}>= <\Sigma^{-}|\bar{u}\,d|\Lambda>
\label{du2}
\eea
The   above relations Eq. (\ref{su1}) and Eq. (\ref{su2}) are
quite general and apply to matrix elements of any  $SU(3)$ octet
 $\Delta S=1$ operator, like the 
$\Delta S=1$ axial vector  current $\bar{u}\,\gamma_{\mu}\,s $ in
hyperon semileptonic decays.

 With   $ <B'|\partial_{\mu}(\bar{u}\,\gamma_{\mu}\,s)|B>$ given by
$ (f_{1})_{B\to B'}(m_{B'} -   m_{B})$, where $(f_{1})_{B\to B'}$
the vector form factor at $q^{2}=0$ momentum transfer in the 
vector current  $<B'|\bar{u}\,\gamma_{\mu}\,s|B>$ matrix element, we have;
\bea
&& [(1/4)(m_{\Xi^{-}} -m_{\Sigma^{0}}) +(3/4)(m_{\Xi^{-}} -m_{\Lambda})]=[(1/4)(m_{\Sigma^{0}}-m_{p}) +(3/4)(m_{\Lambda}-m_{p})]
\label{GMO1}\\
&& [(m_{\Xi^{-}} -m_{\Sigma^{0}}) -(m_{\Xi^{-}} -m_{\Lambda}))=-[(m_{\Sigma^{0}}-m_{p}) -(m_{\Lambda}-m_{p})]
\label{GMO2}
\eea
Eq. (\ref{GMO1})  reproduces the  GMO relation given in  Eq. (\ref{GMO})
mentioned above. Eq. (\ref{GMO2}) reduces to a trivial identity
with  both its l.h.s and r.h.s equal to $-(m_{\Sigma^{0}}-m_{\Lambda})$.
Experimentally, the l.h.s and r.h.s of  Eq. (\ref{GMO1}) is $0.1867\,\rm GeV$
and  $0.1966\,\rm GeV$  respectively, showing a small 
$SU(3)$ symmetry breaking effects, of the order $d=0.05$, the ratio of 
the difference between the l.h.s and r.h.s to the average 
of the two quantities.  One therefore expects a similar amount of
symmetry breaking in hyperon semileptonic decays, appearing  as a
violation of the  axial vector current   GMO relations which are 
obtained easily by making a substitution 
$\bar{u}\,s \to \bar{u}\gamma_{\mu}\gamma_{5}\,s$ in
Eqs. (\ref{su1},\ref{su2}). We have,
\bea
&& <\frac{1}{2}\Sigma^{0} +\frac{\sqrt{3}}{2}\Lambda|\bar{u}\gamma_{\mu}\gamma_{5}\,s|\Xi^{-}>= 
-<p|\bar{u}\gamma_{\mu}\gamma_{5}\,s|\frac{1}{2}\Sigma^{0} +\frac{\sqrt{3}}{2}\Lambda>\label{asu1}\\
&& <\frac{\sqrt{3}}{2}\Sigma^{0} -\frac{1}{2}\Lambda|\bar{u}\gamma_{\mu}\gamma_{5}\,s|\Xi^{-}>= 
<p|\bar{u}\gamma_{\mu}\gamma_{5}\,s|\frac{\sqrt{3}}{2}\Sigma^{0} -\frac{1}{2}\Lambda>
\label{asu2}
\eea
 In terms of   $(g_{1}/f_{1})_{B\to B'}$ the axial vector current to vector
current form factor ratios \cite{Bourquin}, we find,
\bea
&& (1/4)(g_{1}/f_{1})_{\Xi^{-}\to \Sigma^{0}} + (3/4)
(g_{1}/f_{1})_{\Xi^{-}\to \Lambda}= 
(1/4)(g_{1}/f_{1})_{\Sigma^{0}\to p} + (3/4)
(g_{1}/f_{1})_{\Lambda \to p}\label{GMOa}\\
&& (3/4)[(g_{1}/f_{1})_{\Xi^{-}\to \Sigma^{0}} - 
(g_{1}/f_{1})_{\Xi^{-}\to \Lambda}]= 
-(3/4)[(g_{1}/f_{1})_{\Sigma^{0}\to p} - 
(g_{1}/f_{1})_{\Lambda \to p}]\label{GMOb}
\eea

 Since the measured $(g_{1}/f_{1})_{B\to B'}$ contain first 
and second order $SU(3)$ breaking effects ($f_{1}$ has only second order
$SU(3)$ breaking according to the Ademollo-Gatto theorem as mentioned above),
there will be violation of the above relations by first and second
order $SU(3)$ breaking terms, though the violation due to second order
$SU(3)$ breaking could be less important due to possible cancellation
of second order $SU(3)$ breaking effects in $(g_{1}/f_{1})_{B\to B'}$.
Thus  the validity of the above relations would depend essentially 
on  first order $SU(3)$ symmetry breaking effects.

\begin{table}[h]
\begin{tabular}{|c|c|c|c|c|c|}
\hline
\hline
 Decay &$f_{1}$& $(g_{1})_{SU(3)}$& $(g_{1}/f_{1})_{SU(3)+ \rm SB}$&
 $(g_{1}/f_{1})_{\rm exp}$\cite{PDG,Bourquin2} & $d_{B\to B^{'}}$(\rm estimated) \\ 
\hline
$n\to p\ell\bar{\nu}$ &$1$ &$F + D$ &$ F + D$&$1.2694\pm 0.0028 $&$$\\
$\Lambda\to p\ell\bar\nu$&$-\sqrt{3/2}$  &$-\sqrt{3/2}(F +D/3)$ & $F +
 D/3 + d_{\Lambda \to p}$&$0.718\pm 0.015 $&$ -0.015- 0.011 $\\
$\Sigma^-\to n\ell\bar\nu$&$-1$ &$-F + D$&$F -D + d_{\Sigma^{-}\to
 n}$&$-0.340\pm 0.017 $&$ -0.034 $(\rm input) \\
$\Xi^-\to\Lambda^0\ell\bar\nu$ &$\sqrt{3/2}$&$\sqrt{3/2}(F -D/3)$&$F
 -D/3+d_{\Xi^{-}\to \Lambda}$&$0.25\pm 0.05$&$0.053 - 0.023$ \\
$\Xi^0\to\Sigma^+\ell\bar\nu$ &$1$&$F + D $ &$F + D+d_{\Xi^{0}\to
 \Sigma^{+}} $&$1.21\pm 0.05 $&$-0.06$(\rm data)\\
$\Xi^-\to\Sigma^0\ell\bar\nu$ &$1/\sqrt{2}$& $(1/\sqrt{2})(F + D)$ &$F +
 D + d_{\Xi^{-}\to \Sigma^{0}}$&$ $&$ $\\
$\Sigma^-\to\Lambda\ell\bar\nu$ &$0$&$\sqrt{2/3}D $ &$ $&$(g_{1})_{\rm exp}=$&$-0.070 - 0.028$\\
$\Sigma^+\to\Lambda\ell\bar\nu$ &$0$&$\sqrt{2/3}D $ &$ $& $0.587\pm 0.016$&$-0.070 - 0.028$ \\
\hline
\hline
\end{tabular}
\caption{Vector and axial vector current form factors for baryon 
semileptonic decays in the Cabibbo  model and with $SU(3)$ breaking 
term $d_{B\to B^{'}}$ and the measured axial vector to vector form 
factor ratio $g_{1}/f_{1}$, the $SU(3)$ and measured values for
 $(g_{1})_{\Sigma^-\to\Lambda}$. The last column is the estimated
$d_{B\to B^{'}} $.}
\label{table1}
\end{table}

 In the exact $SU(3)$ symmetry limit,  the l.h.s and r.h.s of
 Eq. (\ref{GMOa}) are equal as well as that of Eq. (\ref{GMOb}), given 
by $F$ and $D$, respectively. F is just
 $(g_{1}/f_{1})_{\Sigma^{+}\to \Sigma^{0}}$ and
$-(g_{1}/f_{1})_{\Sigma^{0}\to \Sigma^{-}}$, while $D$ is
$\sqrt{3/2}(g_{1})_{\Sigma^{+}\to \Lambda}$ and $\sqrt{3/2}(g_{1})_{\Lambda\to
  \Sigma^{-}}$ as mentioned earlier. In  the presence of  $SU(3)$
 symmetry breaking, the l.h.s and r.h.s of Eq. (\ref{GMOa}) and 
 Eq. (\ref{GMOb}) differ and are given by,
\bea
&& L_{1} = F + (1/4)\,d_{\Xi^{-}\to \Sigma^{0}} + (3/4)\,d_{\Xi^{-}\to
  \Lambda}, \quad R_{1} = F + (1/4)\,d_{\Sigma^{0}\to p} +
(3/4)\,d_{\Lambda \to p} 
\label{GMOa1}\\
&& L_{2} = D + (3/4)\,(d_{\Xi^{-}\to \Sigma^{0}} -d_{\Xi^{-}\to
  \Lambda}), \quad \quad \  R_{2}= D - (3/4)\,(d_{\Sigma^{0}\to p} -
d_{\Lambda \to p}) 
\label{GMOb1}
\eea

 In our analysis, we define the $SU(3)$
breaking terms with respect to the neutron $\beta$ decay amplitude and
our $D,F$ are pure $SU(3)$-symmetric parameters.

  Since 
$<\Sigma^{0}|\bar{u}\gamma_{\mu}(1-\gamma_{5})\,s|\Xi^{-}>=<\Sigma^{+}|\bar{u}\gamma_{\mu}(1-\gamma_{5})\,s|\Xi^{0}>/\sqrt{2} $,
$<p|\bar{u}\gamma_{\mu}(1-\gamma_{5})\,s|\Sigma^{0}>=<n|\bar{u}\gamma_{\mu}(1 -\gamma_{5})\,s|\Sigma^{-}>/\sqrt{2} $,
 for both vector and axial vector current matrix elements,
 $(g_{1}/f_{1})_{\Xi^{-}\to \Sigma^{0}}=(g_{1}/f_{1})_{\Xi^{0}\to \Sigma^{+}} $
and 
$(g_{1}/f_{1})_{\Sigma^{0}\to p }=(g_{1}/f_{1})_{\Sigma^{-}\to n} $
one can use the measured values $(g_{1}/f_{1})_{\Xi^{0}\to \Sigma^{+}} $ 
and $(g_{1}/f_{1})_{\Sigma^{-}\to n} $ to test $SU(3)$ symmetry in 
hyperon semileptonic decays. 

 The differences $\Delta_{1}= L_{1} - R_{1}$ and $\Delta_{2}= L_{2} - R_{2}$
depend only on the symmetry breaking terms and are  measures of
$SU(3)$ symmetry breaking. We have,
\bea
&&\Delta_{1}=  (1/4)( d_{\Xi^{-}\to \Sigma^{0}}-d_{\Sigma^{0}\to p})+ 
(3/4)(d_{\Xi^{-}\to  \Lambda} - d_{\Lambda \to p})\label{Delta1}\\
&&\Delta_{2}=  (3/4)( d_{\Xi^{-}\to \Sigma^{0}}-d_{\Xi^{-}\to \Lambda})+ 
(3/4)(d_{\Sigma^{0}\to p} - d_{\Lambda \to p})
\label{Delta2}
\eea
 From the measured values in Table
\ref{table1}, we have,
\bea
&&L_{1} =0.490\pm 0.05, \quad R_{1} =0.453\pm 0.015, \quad \Delta_{1}=
 0.036\pm 0.065\label{R1}\\
&&L_{2} =0.720\pm 0.075, \quad R_{2} =0.793\pm 0.024, \quad
 \Delta_{2}=-0.073\pm 0.10 \label{R2}
\eea
showing on average, an  amount of $SU(3)$ breaking  of $4\%$ from
 $\Delta_{1}$ and $10\%$ from
$\Delta_{2}$ (ignoring experimental errors), to be compared with
an amount of $SU(3)$ breaking of $5\%$ in the
  $<B^{'}|\bar{u}s|B>$ matrix element from the GMO mass formula.
For the $\Sigma^-\to\Lambda\ell\bar{\nu}$ decays, the measured value of
$0.719\pm 0.022$ for $\sqrt{3/2}(g_{1})_{\Sigma^-\to\Lambda}$ differs also    
with  $L_{2} $ and $R_{2}$ in Eq. (\ref{R2}),  showing an $SU(3)$ breaking
 effect  of  $11\%$ in $\Sigma^-\to\Lambda\ell\bar{\nu}$ decays. 

 Using the measured 
 $(g_{1}/f_{1})_{n \to p}$ and  $(g_{1}/f_{1})_{\Sigma^{-}\to n}$, we have, 
\be
F =  0.464 -d_{\Sigma^-\to n}/2, \quad D =  0.805 +d_{\Sigma^-\to n}/2
\label{FD0}
\ee
The  $SU(3)$-symmetric fit of Ref. \cite{Gensini} produces an  $SU(3)$
value $(g_{1}/f_{1})_{\Sigma^{-}\to n}=-0.3178$ to be compared with the
 measured value of $-0.340\pm 0.017$. This implies an $SU(3)$ breaking 
of $6.5\%$. This value is comparable with the calculations
 of Ref. \cite{Holstein}
which give a $7.8\%$ $SU(3)$ breaking. Including possible
uncertainties in these values, we shall take  $d_{\Sigma^-\to n}=-0.034$ 
for our determination of the symmetry breaking terms $d_{B\to B^{'}}$
and $D, F$. From Eq. (\ref{FD0}), we find,
\be
F =  0.464 + 017, \quad D =  0.805  - 0.017 
\label{FD}
\ee
Thus  the  symmetry breaking for $(g_{1}/f_{1})_{\Sigma^{-}\to n}$
 make  a rather small contribution to  $F$ and $D$. We note  the
importance of the small value for $d_{\Sigma^{-}\to n}$ used in the
determination of $D,F$ with results close to the values obtained from the
$SU(3)$-symmetric fit of Ref. \cite{Gensini}. More
 precisely, the fit of Ref. \cite{Gensini} looks like a  zeroth order fit
in $SU(3)$ breaking  and ours is an improved
determination of $D, F$ with $SU(3)$ symmetry breaking removed
according to Eq. (\ref{FD}). From the data  and the above
determined  values for $F$ and $D$, we now determine  the   $SU(3)$
breaking terms in hyperon semileptonic decays, using the above value 
 $d_{\Sigma^{-}\to n}=- 0.034$ and ignoring the experimental
 error of $\pm 0.05$ in the measured 
 $(g_{1}/f_{1})_{\Xi^0 \to \Sigma^{+}} $ and 
$(g_{1}/f_{1})_{\Xi^{-}\to \Lambda} $. We have,
\bea
&&  d_{\Xi^0 \to \Sigma^{+}}= -0.06, \quad \ \ \quad \ \quad  d_{\Xi^{-}\to \Lambda}= 0.053 - 0.023    \nonumber \\
&& d_{\Lambda\to p}= -0.015 - 0.011, \quad  d_{\Sigma^- \to \Lambda}=-0.070 - 0.028.
\label{SB1}
\eea
as  shown in the last column of Table \ref{table1}. Though the symmetry
breaking in $(g_{1}/f_{1})_{\Xi^{-}\to \Lambda} $ is somewhat large ($20\%$),
 the  experimental error of the measured value is also large ($\pm 0.05$), 
one would need a better measurement  for a more accurate estimate of the
symmetry breaking for  $(g_{1}/f_{1})_{\Xi^{-}\to \Lambda} $. In fact, with
$F$ and $D$ obtained in Eq. (\ref{FD}), in the absence of a large symmetry 
breaking term $d_{\Xi^{-}\to \Lambda}$, $(g_{1}/f_{1})_{\Xi^{-}\to \Lambda} $
would be close to $0.20$, a value at present not excluded by experiments.

In the above analysis, we consider  $D, F$  the usual parameters 
of the  $SU(3)$-symmetric part of $(g_{1})_{B\to B^{'}}$ 
 with $(g_{1}/f_{1})_{n\to p}= F + D$ and $d_{B\to B^{'}} $ are 
$SU(3)$ breaking terms including  first and second order $SU(3)$ 
breaking due to  the $s$-quark mass and   contribute to the 
deviation of $ L_{1}, R_{1} $ and $ L_{2}, R_{2} $ from the
$SU(3)$-symmetric  $F$ and $D$ , respectively. If one writes the first order
$SU(3)$ breaking terms in  $d_{B\to B^{'}}$ as the induced terms
produced by a term transforming as the  $8$-component of an $SU(3)$ 
octet \cite{Ademollo}~:
\bea
    \mathcal{L}_{\rm SB} &=&  a_0~{\rm Tr}(\bar{B}B\lambda_i)+
b_0~{\rm Tr}(\bar{B}\lambda_iB)+    a~{\rm Tr}(\bar{B}B\{\lambda_i, \lambda_8\})+   b~{\rm Tr}(\bar{B}\{\lambda_i, \lambda_8\}B)   \nonumber\\
   &&+c~[{\rm Tr}(\bar{B}\lambda_iB\lambda_8)-
   {\rm Tr}(\bar{B}\lambda_8 B\lambda_i)]
   +g~{\rm Tr}(\bar{B}B){\rm Tr}(\lambda_i\lambda_8) \nonumber \\
   &&+h~[{\rm Tr}(\bar{B}\lambda_i){\rm Tr}(B\lambda_8)+
   {\rm Tr}(\bar{B}\lambda_8){\rm Tr}(B\lambda_i)]~,
\label{SBA}
\eea
then  the   $SU(3)$ breaking terms from
 the $\underline{8}$  representations in the above expression  will
 not produce a violation of the relations  Eq. (\ref{GMOa}) and 
Eq. (\ref{GMOb}), like the $SU(3)$-symmetric  $D, F$ terms. The $c$-terms 
 are from the   $\underline{10}$ and $\underline{10}^{*}$ 
representation and the $h$-terms  are from the $\underline{27}$ 
representation. These  terms will produce violation of the relations
 Eq. (\ref{GMOa}) and Eq. (\ref{GMOb})  and  provide clear 
evidence for $SU(3)$ breaking in hyperon semileptonic decays. For
 example,  in  the analysis  of Ref. \cite{Yamanishi}, the $a,b$ terms in
 Eq. (\ref{SBA}) and in  Eqs. (8a-8i) of the paper 
could be absorbed into $a_{0}=D-F$ and $b_{0}=D+F$ terms and thus do
not contribute to $\Delta_{1}$ and $\Delta_{2}$  . Assuming
 no isospin breaking, as in Ref. \cite{Ademollo},  by putting
 $\alpha=0, \beta=1$ in the expressions for $(g_{1}/f_{1})_{B\to B^{'}} $
in Eqs. (8a-8i) of Ref. \cite{Yamanishi}, we have,
\be
\Delta_{1}= h, \qquad \Delta_{2}= 3\,c 
\label{Deltac}
\ee
which allow a determination of $h$ and $c$ from the experimental values for
$\Delta_{1}$ and  $\Delta_{2}$. Note that in the notation of 
Ref. \cite{Holstein}, $H_{3}=3\,c$ and $H_{4}= h$, we have:
\be
\Delta_{1}= H_{4} , \qquad \ \Delta_{2}= H_{3}
\label{DeltaH}
\ee
 From   Eq. (\ref{R1}) and Eq. (\ref{R2}), we find,
\be
c = -0.024 \pm 0.04,\quad  H_{3}= -0.073\pm 0.10 , \quad \ h = H_{4} = 0.036 \pm 0.065
\label{Deltad}
\ee
to be compared  with the corresponding values $H_{3}=-0.006$ and
 $H_{4}=0.037$, given in Ref. \cite{Holstein}. But there is a problem
with this model. If we assume that there is no $SU(3)$ breaking 
in $(g_{1}/f_{1})_{n\to p}$, we would have $b=c$ in the expression for
$(g_{1})_{n\to p}$ in Eq. (8a) of Ref. \cite{Yamanishi}, this  implies
that there would be  no $SU(3)$ breaking in 
 $(g_{1}/f_{1})_{\Xi^{-}\to \Sigma^{0}}$ in contradiction with experiments.
We note also that $d_{\Sigma^{+}\to \Lambda}$ would be large and positive,
in contradiction with the value obtained from the $SU(3)$-symmetric fit
of Ref. \cite{Gensini} and our result shown in the Table \ref{table1}.

\section{Conclusion}

In conclusion, we have shown that the GMO  relations for the baryon 
mass difference is quite general and can be derived for the 
axial vector current matrix elements in hyperon semileptonic decays.
With  these  GMO type relations, we present  evidence of an
 $SU(3)$ breaking, similar to that in the
baryon mass difference. We then  give an estimate for the $SU(3)$-symmetric
$F$ and $D$ terms as well as symmetry breaking terms using the measured  
axial vector form factors. The small symmetry breaking effect we find also
confirms the success of the Cabibbo model for hyperon semileptonic decays.
Finally, these GMO relations could be used as experimental constraints
 on the $SU(3)$ symmetry breaking terms in theoretical calculations.


\begin{thebibliography}{99}

\bibitem{Donoghue} J. F. Donoghue, E. Golowich, and B. R. Holstein, 
{\em Dynamics of the Standard Model},\\ Cambridge University Press, Cambridge,
(1992).


\bibitem{Cabibbo} N. Cabibbo,  Phys.\ Rev.\ Lett {\bf 10}, 531 (1963).

\bibitem{Marshak} R. E. Marshak, Riazuddin and C. P. Ryan, {\em Theory of
  Weak Interactions in Particle Physics},\\ John Wiley, New York(1969).

\bibitem{Ademollo} M. Ademollo and R. Gatto,  Phys.\ Rev.\ Lett {\bf 13}, 
264 (1963).


\bibitem{PDG} Particle Data Group, J. Beringer {\em et al},  Phys.\ Rev.\
  D {\bf 86}, 010001 (2012).

\bibitem{Ratcliffe}  P.~G.~Ratcliffe,  Phys. Lett. B {\bf365}, 383 (1996).

\bibitem{Song} X.~Song, P.~K.~Kabir, and J.~S.~McCarthy,  Phys.\ Rev.\ D
{\bf 54}, 2108 (1996).

\bibitem{Yamanishi} T. Yamanishi, Phys.\ Rev.\ D {\bf 76}, 014006 (2007).

\bibitem{Bourquin} M. Bourquin {\em et al.}, Z. Phys. C {\bf 21}, 27 (1983).

\bibitem{Bourquin2} M. Bourquin {\em et al.}, Z. Phys. C {\bf 12}, 307 (1982).

\bibitem{Gensini} P. M. Gensini and G. Violini, Talk presented at
 {\em the 5th International Symposium on\\ Meson-Nucleon Physics 
and the Structure of the Nucleon}, Boulder, Colorado, 6-10 Sept. 1993,\\
Edited by B.M.K. Nefkens, M. Clajus. Pi N Newslett. (1993) Nos. 8, 9;\\
arXiv:hep-ph/9311270 (1993).


\bibitem{Holstein} A. Faessler, T. Gutsche, B. R. Holstein, M. A. Ivanov,
J. G. K${\ddot o}$rner, \\ and V. E. Liubovitskij, Phys.\ Rev.\ D {\bf 78}, 
094005 (2008).

\end{thebibliography}
\end{document}